\documentclass[twocolumn,showpacs,preprintnumbers,amsmath,amssymb]{revtex4}
\usepackage[dvips]{graphicx}

\begin{document}




\preprint{}

\title{
Single Transverse-Spin Asymmetry in Very
Forward and Very Backward Neutral Particle Production 
for Polarized Proton Collisions 
at $\mathbf{\sqrt{s} =}$ 200~GeV}

\author{Y.~Fukao,$^{3,7,}\footnotemark[1]$
        M.~Togawa,$^{3,6}$
        A.~Bazilevsky,$^{7,}\footnotemark[2]$
        L.~C.~Bland,$^{1}$
        A.~Bogdanov,$^{5}$
        G.~Bunce,$^{1,7}$
        A.~Deshpande,$^{7,}\footnotemark[3]$
        H.~En'yo,$^{6,7}$
        B.~D.~Fox,$^{7,}\footnotemark[4]$
        Y.~Goto,$^{6,7}$
        J.~S.~Haggerty,$^{1}$
        K.~Imai,$^{3,6}$
        W.~Lenz,$^{1}$
        D.~von~Lintig,$^{1}$
        M.~X.~Liu,$^{4}$
        Y.~I.~Makdisi,$^{1}$
        R.~Muto,$^{3,6}$
        S.~B.~Nurushev,$^{2}$
        E.~Pascuzzi,$^{7}$
        M.~L.~Purschke,$^{1}$
        N.~Saito,$^{3,6,7,}\footnotemark[5]$
        F.~Sakuma,$^{3,6}$
        S.~P.~Stoll,$^{1}$
        K.~Tanida,$^{6,}\footnotemark[6]$
        J.~Tojo,$^{3,6,}\footnotemark[5]$
        Y.~Watanabe,$^{6,7}$
    and C.~L.~Woody$^{1}$
}

\affiliation{
$^{1}$\mbox{Brookhaven National Laboratory,~Upton,~NY~11973-5000,~USA}\\
$^{2}$\mbox{Institute for High Energy Physics,~Protvino,~Russia}\\
$^{3}$\mbox{Kyoto University,~Kyoto~606-8502,~Japan}\\
$^{4}$\mbox{Los Alamos National Laboratory,~Los~Alamos,~NM~87545,~USA}\\
$^{5}$\mbox{Moscow Engineering Physics Institute,~State~University~of~Russia,~Russia}\\
$^{6}$\mbox{RIKEN,~Wako,~Saitama~351-0198,~Japan}\\
$^{7}$\mbox{RIKEN-BNL~Research~Center,~Brookhaven~National~Laboratory,~Upton,~NY~11973-5000,~USA}
}

\date{\today}

\begin{abstract}

In the 2001-2002 running period of the Relativistic Heavy Ion Collider (RHIC), transversely 
polarized protons were accelerated to 100~GeV for the first time, with
collisions at $\sqrt{s}$ = 200~GeV.  We present results from this run for single transverse 
spin asymmetries for inclusive production of neutral pions, photons and neutrons 
of the energy region 20 -- 100 GeV for forward and backward production for angles between
0.3 mrad and 2.2 mrad with respect to the polarized proton direction.  
An asymmetry of $A_N = (-0.090 \pm 0.006 \pm 0.009) \times (1.0^{+0.47}_{-0.24})$ was observed 
for forward neutron production, where the errors are statistical and systematic, and the scale error
is from the beam polarization uncertainty.
The forward photon and $\pi^0$, and backward neutron, photon, and $\pi^0$
asymmetries were consistent with zero.  The large neutron asymmetry indicates a strong interference between a
spin-flip amplitude, such as one pion exchange which dominates lower energy neutron production,
and remaining spin non-flip amplitudes such as Reggeon exchange.

\end{abstract}


\pacs{Valid PACS appear here}
\maketitle

\renewcommand{\thefootnote}{\fnsymbol{footnote}}
\footnotetext[1]{
Present address: RIKEN,~Wako,~Saitama~351-0198,~Japan
}
\footnotetext[2]{
Present address: Brookhaven National Laboratory,~Upton,~NY~11973-5000,~USA
}
\footnotetext[3]{
Present address: Stony Brook University, SUNY, Stony Brook, NY 11794, USA
}
\footnotetext[4]{
Present address: Pennsylvania State University, University Park, PA 16802, USA
}
\footnotetext[5]{
Present address: KEK, High Energy Accelerator Research Organization, Tsukuba-shi, Ibaraki-ken 305-0801, Japan
}
\footnotetext[6]{
Present address: Kyoto University,~Kyoto~606-8502,~Japan
}

Polarized proton beams were accelerated to 100 GeV for the first time in the 2001-2002 running
period of the Relativistic Heavy Ion Collider (RHIC).  We present results from an experiment at the 12 o'clock
intersection point (IP12), for the single-spin asymmetry $A_N$ for collisions of transversely
polarized protons with unpolarized protons, producing very forward 
(and very backward) neutral particles.  The goal of the
measurement was to search for a production process with a significant asymmetry that could be used
as a local polarimeter at the experiments, to monitor the polarization direction in collision at RHIC. 
The very forward region was chosen
due to accessibility with the existing RHIC detectors.  The experiment was designed to search for
photon and $\pi^0$ asymmetries, based on the observation of a large pion asymmetry for 
$\sqrt{s}$ = 19.4~GeV.\cite{E704}  However, a large neutron asymmetry was observed, which is reported here.


The measured cross sections for inclusive production of neutrons at large $x_F$
for unpolarized proton-proton collisions are consistent with one pion exchange
(OPE) model predictions from $\sqrt{s}$ = 7 to 64~GeV.\cite{OPE,Flauger76,Blobel78,D'Alesio99}
The model has been used to describe exclusive diffractive production of neutrons \cite{Berger75},
production near the exclusive limit at large $x_F$, and inclusive
neutron and proton production away from this limit (see \cite{OPE}).  More recently,
with interest in HERA $e$-$p$ collider studies of proton fragmentation, OPE is used
to describe proton and photon induced production of neutrons which may probe the pion
structure function at small $x$.\cite{Kopeliovich96,D'Alesio99} For polarized phenomena,
Soffer and T\"{o}rnqvist \cite{Soffer} applied the OPE model to describe the polarization
of inclusive $\Lambda$ production in $p$-$p$ collisions.\cite{Bunce} A single transverse
spin asymmetry, the analyzing power $A_N$, must arise from an interference of
spin flip and spin non-flip amplitudes.  The OPE amplitude is
fully spin flip; a spin non-flip amplitude contribution can be described 
in Regge theory as due to Reggeon and Pomeron exchange.\cite{regge} To the extent
that the cross section is described well by spin-flip OPE,
$A_N$ is a sensitive test of this picture, and is sensitive to a small
admixture of a non-spin flip contribution at the amplitude level.
  
The analyzing power is the azimuthal modulation of the cross section, or left-right
asymmetry of a vertically polarized beam, of produced particles relative to the spin direction 
of a transversely polarized beam on an unpolarized target.
The measurement consists of observing the azimuthal dependence of the production of 
identified particles, normalized by an independent measurement of the beam polarization.   

The RHIC polarized proton beams were vertically polarized,
and bunched, with a bunch length of $\sim$3~ns, with the bunches separated by 212~ns.  The 55 bunches in one
ring (the Blue ring) had alternating spin directions in successive bunches, and the other
ring (the Yellow ring) had alternating spin directions for successive pairs of bunches.

RHIC timing signals identified the bunches and their polarization sign.
This provided collisions
between protons with all combinations of vertical spin orientation, at the same time.  Asymmetries
could be measured for either beam polarization, corresponding to very forward or very backward production,
by averaging over the polarization directions of the opposing beam.  After averaging over the spin directions
for one beam, and over many fills of RHIC, the residual beam polarization was less than 1\% of the RHIC
beam polarization.  A typical
fill of RHIC for this first run had 3 $\times$ 10$^{10}$ protons in each bunch.
With $\beta^*$ = 3 m at the IP12 collision location, the typical 
luminosity was $L$ = $1 \times 10^{30}$ cm$^{-2}$s$^{-1}$
and the average beam divergence was 0.1 mrad rms.

The polarization of each beam was
measured by scattering from ultra-thin carbon targets that were inserted into the beams for
the measurements, at a location where the beams are separated.\cite{jinnouchi}  Measurements took less than 
one minute, giving a polarization sensitivity
of about 3\%, and were typically taken every two hours of a six hour store.
The measured beam polarizations for the
2001-2002 run were $P_{\mbox{Blue}} = (0.11 \pm 0.002 \pm 0.02) \times (1.0 \pm 0.32)$, and
$P_{\mbox{Yellow}} = (0.16 \pm 0.002 \pm 0.02) \times (1.0 \pm 0.32)$.
The first uncertainty is statistical, the second the systematic on the asymmetry
measurement, and the scale error is from the calibration of the polarimeter analyzing power.\cite{Tojo,trueman}

\begin{figure}
\includegraphics[width=1.0\linewidth]{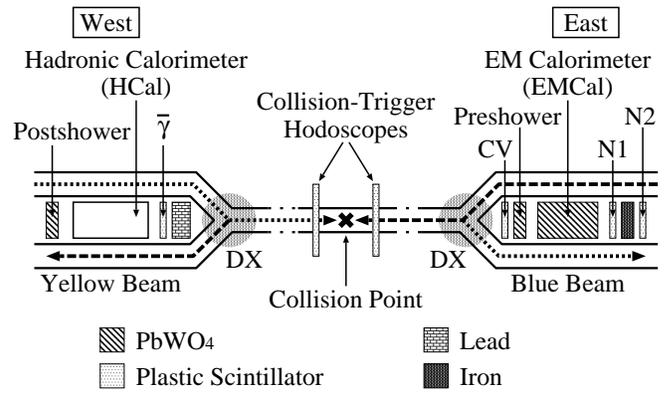}
\caption{A plan view of the experiment at IP12, not to scale. Shown are the principal components
of the experiment: the beam-beam collision trigger hodoscopes, the electromagnetic calorimeter (east)
facing the Blue beam, and the hadronic calorimeter (west) facing the Yellow beam.}
\label{fig:layout}
\end{figure}

The experimental apparatus at IP12 consisted of a pair of scintillator hodoscopes 
placed forward and backward of the collision point
that identified proton-proton collisions, and detectors placed in the very forward directions,
centered on a production angle of 0$^{\circ}$ downstream of the RHIC DX dipole magnets, which removed produced charged particles
(Fig.~\ref{fig:layout}).
The collision-trigger hodoscopes (beam-beam counters) 
were located $\pm$1.85 m from the IP center.
The hodoscopes were formed with four sets of rectangular scintillators, with full azimuthal coverage, and pseudorapidity acceptance
$|\eta |$ = 2.2 to 3.9 in the vertical and horizontal directions from the beams.
The time resolution of the hodoscopes gave a vertex resolution of 23 cm.  The vertex distribution
had a 54~cm rms, and non-collision events within the collision trigger were estimated to be 3\%.

The east detector consisted of a scintillation counter
charged particle veto (dimensions 10.5~cm $\times$ 25~cm $\times$ 0.6~cm, CV in Fig.~\ref{fig:layout}),
a preshower electromagnetic calorimeter, and a 60-element array of
PbWO$_4$ crystal electromagnetic calorimeter (EMCal).  These were followed by scintillation counters on either side of
an absorber to observe shower leakage.  
The preshower and EMCal arrays were built from PbWO$_4$ crystals $2.0\!\times\! 2.0\!\times 20.0$
~cm$^3$, coupled to 3/4-inch photomultiplier tubes (HPK R4125) via silicon cookies through filters of 1/10 light reduction.
The preshower consisted of five
crystals forming a horizontal hodoscope, 10 cm (horizontal) $\times$ 20 cm (vertical) $\times$ 2 cm (deep).
The EMCal was a 5 (horizontal) $\times$ 12 (vertical) crystal array, 20 cm (22 radiation lengths) deep.  
The preshower and EMCal were calibrated in the electron beam of the Final Focus Test Beam Facility
at SLAC.
The energy resolution was $\Delta E/E \simeq$ 10\%/$\sqrt{E\mbox{(GeV)}}$ and the
position resolution was 0.1~cm, independent of the hit position.  The energy resolution was confirmed
by the width of the observed $\pi^0$ peak during the experiment. 
For hadronic showers, the position resolution
was estimated to be 0.5~cm using the simulation program {\sc Geant3}\cite{GEANT}, which had been
tuned to match the EMCal response to the electrons.
However, the EMCal was 1 interaction length with an estimated deposited
energy fraction for neutrons of $E_{dep} : E_n \sim 1 : 3$ with a $>$ 100\% energy resolution.
The EMCal was followed by two scintillators that covered
the area of the EMCal (dimensions 10~cm $\times$ 24~cm $\times$ 0.6~cm), with a 2.8~cm thick iron block in between,
N1 and N2 in Fig.~\ref{fig:layout}. 

Data were taken with the primary trigger requirement of a collision, with $\geq$1 minimum ionizing
particles in each beam-beam hodoscope, within a 5~ns collision window,
and with a minimum energy deposited in
the EMCal, $>$5~GeV.  Analog to digital converters recorded the energies
for each calorimeter block, time to digital converters recorded the summed signals from the beam-beam
counters, and each counter was latched.
The polarization direction information (up or down) was provided by the RHIC
accelerator, and the recorded data were sorted by the polarization directions
of the Blue and Yellow beams.
The trigger rate was 100--300~Hz, and the average live time of the
experiment was 75\%. 70 million events were collected.

Events were reconstructed in the EMCal by combining clusters of hits to provide a total observed
energy, and with positions of the showers obtained from the energy-weighted average of the positions of
the hit blocks.
 
\begin{figure}
\includegraphics[width=1.0\linewidth]{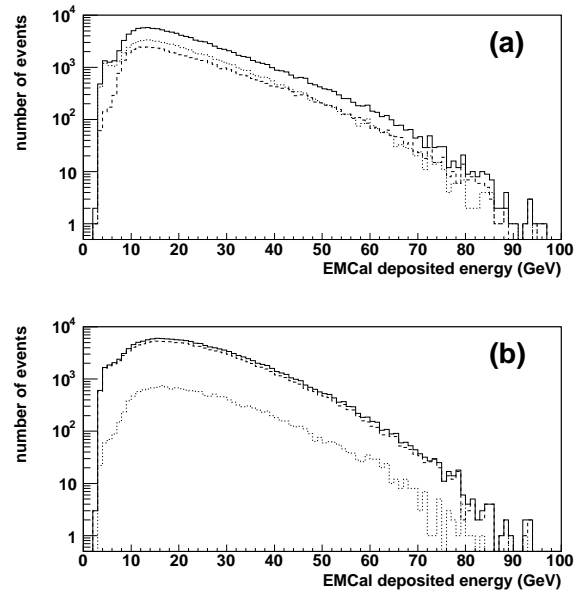}
\caption{Panel (a) shows the deposited energy distribution in
the EMCal for events used to select the photon sample, for events with no charged veto, and no postshower
energy.  The solid curve displays all events, the dashed curve 
displays events with preshower energy (this is the $\gamma$-ID sample),
and the dotted curve shows events with no preshower energy.  Panel (b) shows the energy
distribution for the events used to select neutrons, for events with no charged veto, and with
postshower activity.  The solid curve is for all events, the dashed curve is for events with no
preshower energy ($n$-ID sample), and the dotted curve is for events with
preshower energy.}
\label{fig:particleid}
\end{figure}

Photon samples and neutron samples were identified using the observed preshower and postshower energy.  The photon samples
were identified as events with no charge veto hit, preshower energy higher than 15~MeV, and no postshower activity (``$\gamma$-ID'').
The neutron-sample identification required no charge veto hit, no preshower energy,
and postshower activity (``$n$-ID'').
Both are triggered by the collision requirement and energy in the
EMCal, $>$5~GeV.  Fig.~\ref{fig:particleid} shows the
distribution of energy
deposited in the EMCal for the
photon-sample (a) and neutron-sample (b) selection requirements.
Note that for Fig.~\ref{fig:particleid} (b), the spectrum shape is
predominantly the result of the poor energy resolution of the EMCal for neutrons.

The fraction of photons and neutrons
in each sample was obtained by comparing the fractions of events in the samples with and
without preshower activity.  These fractions can be calculated based on the number of
real photons and neutrons in each sample, using the preshower conversion probabilities
for photons and neutrons.
For the fraction of neutron conversions in the preshower that have post shower
activity, to the fraction that convert in the EMCal with
observed energy in the post-shower, we have assigned the relative probability
of $0.5 \pm 0.5$.
The fraction of neutrons with energy $>$20~GeV in the $\gamma$-ID sample is $0.18 \pm 0.07$ (the purity of photons is $P_\gamma = 0.82 \pm 0.07$);
the fraction of photons in the $n$-ID sample is $0.006 \pm 0.005$ (the purity of neutrons is $P_n = 0.994 \pm 0.005$). 

\begin{figure}
\includegraphics[width=1.0\linewidth]{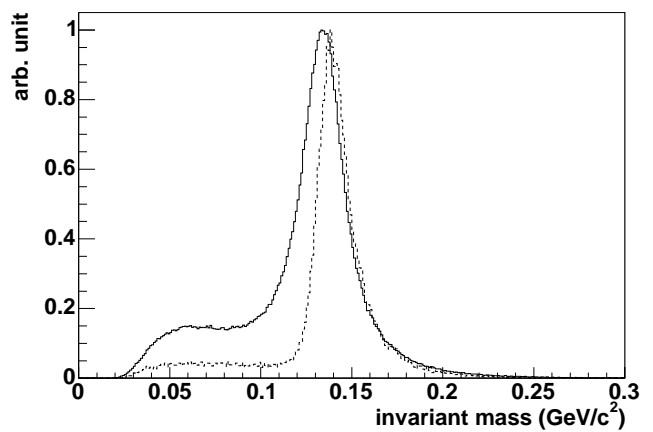}
\caption{Invariant mass of pairs of energy clusters in the EMCal,
for 444K events with no additional selection requirements (solid) and for
35K events with photon identification (dashed).}
\label{fig:pi0mass}
\end{figure}

Fig.~\ref{fig:pi0mass} shows the invariant mass for events with two clusters
of energy in the EMCal, for no additional requirement for photon
identification, and for the $\gamma$-ID sample.
The $\pi^0$ peak mass shifts lower by 5 MeV when no photon identification is
required due to energy lost in photon conversion upstream; both the mass shift
and reduction in number of events are understood from simulation using the
upstream geometry, in particular conversion in a 2.1~cm thick stainless steel
plate just downstream of the DX magnet shown in Fig.~\ref{fig:layout}.
The $\pi^0$ peak center for events with no photon identification was
used for an absolute energy calibration and tower-by-tower final calibration.
To measure the analyzing power for $\pi^0$ events, we use the larger data
sample without photon identification, selecting events within $\pm$20~MeV of
the peak center, with 21\%~$\pm$~3.8\% background, estimating the background
using a Gaussian + polynomial fit.

To measure a left-right asymmetry, it is necessary to eliminate 0$^\circ$ production.
We required the observed shower center to have radii 5 -- 40 mm from the center of the detector, or production angles 0.3 -- 2.2 mrad.
The EMCal hits were assigned coordinates (r,$\phi$) about the 0$^\circ$ reference axis, and binned in $\phi$ counterclockwise, with $\phi$ = 0
along the vertical axis.
A vertical beam polarization would give a left-right asymmetry for $\phi$ = $\pi$/2, and a 
radial beam polarization would give an asymmetry for $\phi$ = 0 and $\phi$ = $\pi$.

To calculate the asymmetry, we used a
method of analysis that reduces systematic errors from detector and luminosity asymmetries.\cite{spinka}  We
defined the geometric mean for the number of observed events with beam proton spin up, producing particles
to azimuthal angle $\phi$, $N_{\uparrow ,\phi}$, and the number of events with beam spin down, producing
particles to azimuth $\phi +\pi$, $N_{\downarrow ,\phi +\pi}$,
$\sqrt{N_{\uparrow ,\phi} N_{\downarrow ,\phi+\pi}}$.
A corresponding geometric mean was defined
for events with spin down, to angle $\phi$, and events with spin up, to angle $\phi +\pi$.  The asymmetry
is the difference of the geometric means, divided by the sum, and normalized by the measured beam
polarization $P_B$: 

\begin{equation}
{\cal A}(\phi)
= \frac{1}{P_B} \frac{\sqrt{N_{\uparrow ,\phi} N_{\downarrow ,\phi+\pi}}
                    - \sqrt{N_{\uparrow ,\phi+\pi} N_{\downarrow ,\phi}}}
                     {\sqrt{N_{\uparrow ,\phi} N_{\downarrow ,\phi+\pi}}
                    + \sqrt{N_{\uparrow ,\phi+\pi} N_{\downarrow ,\phi}}}
\label{eqn:asym}
\end{equation}

\begin{figure}
\includegraphics[width=1.0\linewidth]{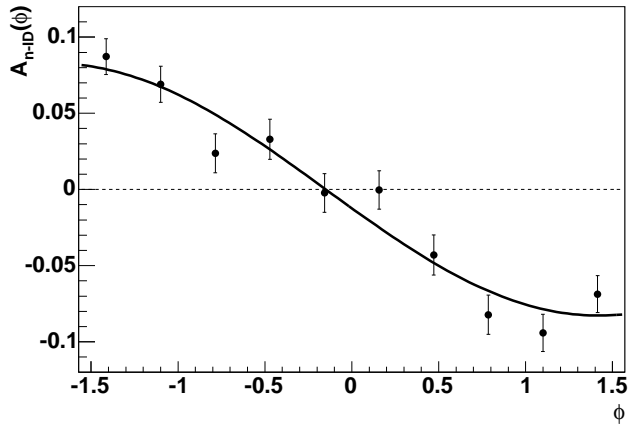}
\caption{Azimuthal dependence of asymmetry for the $n$-ID sample produced
forward with respect to the polarized proton direction, based on the east detector.
The error bars are statistical.}
\label{fig:Aneutron}
\end{figure}

We define $A_N$ as the amplitude of the asymmetry modulation:
\begin{equation}
{\cal A}(\phi ) = A_N \sin(\phi-\phi_0),
\end{equation}
where $\phi_0$ allows a deviation of the polarization direction from $\phi = 0$.
Fig.~\ref{fig:Aneutron} shows the azimuthal dependence of asymmetry for the $n$-ID sample for forward production
with the Blue beam polarized and summing over the spin states of the Yellow beam. 
A significant non-zero asymmetry $A_N^{\mbox{$n$-ID}}$ was observed with $\phi_0 = -0.15 \pm 0.07$ and a reduced $\chi^2 = 1.5$.
Similarly the asymmetries $A_N^{\mbox{$n$-ID}}$ for the $n$-ID sample for backward production (Yellow beam polarized, averaging over
spins of the Blue beam) and $A_N^{\mbox{$\gamma$-ID}}$ for the
$\gamma$-ID sample for forward and backward production were obtained.
Polarized Blue beam results used $\phi_0 = -0.15$, and polarized
Yellow beam results used $\phi_0 = 0$.

The asymmetries of neutrons ($A_N^n$) and photons ($A_N^{\gamma}$) were calculated by:
\begin{equation}
\begin{array}{rcl}
A_N^{\mbox{$n$-ID}}      & = & (1 - P_n) \cdot A_N^{\gamma} + P_n \cdot A_N^n \\
A_N^{\mbox{$\gamma$-ID}} & = & (1 - P_{\gamma}) \cdot A_N^n + P_{\gamma} \cdot A_N^{\gamma}
\end{array}
\end{equation}
where $P_n$ and $P_{\gamma}$ are the purity of neutrons in the $n$-ID sample and that of photons in the $\gamma$-ID sample respectively.

To confirm this neutron asymmetry signal during the run, we added a partial hadronic calorimeter (HCal) facing the Yellow beam, the west detector 
in Fig.~\ref{fig:layout}.  The setup
included a photon veto ($\bar{\gamma}$ in Fig.~\ref{fig:layout}), consisting of a 5-cm thick lead block followed by a scintillator,
the hadron calorimeter, and five PbWO$_4$ crystals that were set up as a horizontal hodoscope to identify left vs. right
production.
The HCal had a transverse dimension of
10~cm $\times$ 10~cm, and a length of 2 interaction lengths.  The PbWO$_4$ crystals were the same dimension
as used for the east detector.  The position resolution of the shower center was estimated
to be 3 to 4~cm, from a {\sc Geant3} simulation.\cite{GEANT}  The asymmetry was obtained by comparing 
the left and right scattering 
only (the azimuthal dependence was not available
with this setup), and was corrected by the factor 0.69 for the integration over the azimuth, due to
the azimuthal dependence of the analyzing power for vertical beam polarization.  For this
measurement, neutrons with a shower maximum in the central crystal were not included, and a HCal
energy of 28~GeV or higher in the trigger and 30~GeV in the offline analysis was required.
The energy resolution of the HCal was 40--50\% for energy greater than 20 GeV.
Thus, the neutron event selection was matched as far as possible for
the HCal and EMCal
setups.  The measured forward asymmetry with the Yellow beam polarized, and summing over the spin states of the Blue beam,
was $A_N = -0.135 \pm 0.018$, a significant and large asymmetry, and
consistent with the result from the east detector at the 2-$\sigma$ level after including systematic uncertainties.


The HCal was a prototype of the Zero Degree Calorimeters (ZDC) used at RHIC, one third of the length of the RHIC ZDCs.\cite{ZDC}
With only two interaction lengths. the measured energy can range from a small
fraction of the actual neutron energy for conversions near the downstream end of the HCal,
to the full neutron energy.  The neutron asymmetry reported here has been used
in subsequent RHIC runs by the PHENIX experiment to set the proton spin direction
at collision.  A neutron energy spectrum has been reported \cite{Togawa}, which is
consistent with the ISR results \cite{Flauger76}, in particular showing a peak in the
energy spectrum at about $x_F$ = 0.8.  For the experiment reported here, the range
in $p_T$ for the neutrons can only be estimated from the limits of the acceptance,
and the minimum energy requirement, which give 0.01 GeV/$c$ $< p_T <$ 0.22 GeV/$c$, for the
limits 30 GeV $< E_n <$ 100 GeV and 0.3 mrad $< \theta_n <$ 2.2 mrad.  Greatly improved
information on the kinematical variables will be forthcoming from the PHENIX
experiment.


For both calorimeters, the identified neutron sample may include $K^0_L$. 
At ISR energies, the $K^0_L$ fraction to neutrons for a similar kinematical region was estimated
to be 3-4\%, from observed charged kaon samples.\cite{Flauger76}  We have included no correction
for $K^0_L$ background in the neutron results.


Table \ref{tbl:asym} gives the corrected asymmetries from the east detector for neutron, photon, and $\pi^0$ forward production
and for backward production.
The systematic uncertainties include contributions from scattered beam background, particle misidentification, and
from smearing of the hit position in the EMCal.
Scattered beam background refers to an observed higher rate in the EMCal towers near the beam pipe of the outgoing
RHIC beam.
Whether or not the background has a spin dependence, it affects the azimuthal dependence of Eq.(\ref{eqn:asym}).
Its contribution to the systematic uncertainty was estimated from the reduced $\chi^2 = 1.5$ of the azimuthal fit to
${\cal A}_{\mbox{$n$-ID}}(\phi)$, where we have assigned a systematic uncertainty of
$\sqrt{0.5} \times \delta A_N(\mbox{statistical})$.
The scattered beam background uncertainties for the photon and $\pi^0$ are negligible.
The particle identification uncertainties were presented earlier.
A hit smearing correction of 0.92~$\pm$~0.08 was evaluated using a {\sc Geant3} simulation of the EMCal which described
the test beam result for position resolution.
Finally, the beam polarization uncertainty, presented earlier, scales both the asymmetry and uncertainties presented in
Table \ref{tbl:asym}, preserving the measurement significance.
The scale factor is $(1.0^{+0.47}_{-0.24})$.


All asymmetries use events with energies 20 -- 100 GeV in the EMCal and production angles 0.3 -- 2.2 mrad.
We observe asymmetries consistent with zero for photon and $\pi^0$, for both forward and backward
production, and for neutron backward production, and a significant non-zero asymmetry for
forward neutron production.  We have also measured the asymmetry for the photon sample obtained from
conversions near the DX magnet, where the neutron contamination is estimated to be less than 12\%.
This asymmetry is $0.016 \pm 0.020$, consistent with zero.

\begin{figure}
\includegraphics[width=1.0\linewidth]{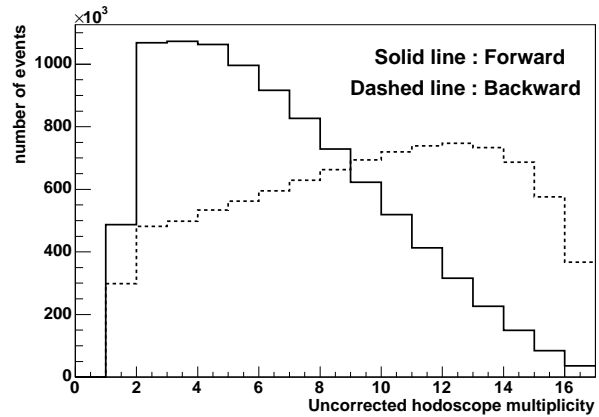}
\caption{Uncorrected multiplicity observed by the beam-beam counters
for the $n$-ID sample of the EMCal data with deposited energy $>$20~GeV.
The solid line shows that for the forward counter (east)
and the dotted line shows that for the backward counter (west).}
\label{fig:hmulti}
\end{figure}

In Fig.~\ref{fig:hmulti} we present the uncorrected multiplicity
observed by the beam-beam counters for the $n$-ID sample of the EMCal data with deposited energy $>$20~GeV, where the trigger
required $\geq$ 1 hit both forward and backward.
The vertical and horizontal slats for each hodoscope overlap over 67\% 
of the area, so that a correct estimate of the multiplicity is $\sim$
(uncorrected multiplicity) / 1.67.
As seen in the figure, the multiplicity for the forward beam counter, 
in the direction of the neutron, is low, $\langle$multiplicity
(forward)$\rangle$ $\sim$ 2.
The multiplicity in the backward beam counter is large,
$\langle$multiplicity (backward)$\rangle$ $\sim$ 7.
Therefore, we observe a clear separation of beam and target
fragmentation multiplicity, with a large asymmetry for neutrons
produced forward from the polarized beam, in the direction of low
multiplicity.
The backward multiplicity is large.
This pattern was confirmed with the HCal data, where the 
forward and backward beam counters are reversed.
The low forward multiplicity may be consistent with a simple exchange
process producing the forward neutron, $p \rightarrow n$, although the
acceptance of the forward beam counter, with 43 mrad $< \theta <$ 297
mrad, covers much larger angles than the opening angle of, for example, $\Delta^+ \rightarrow n + \pi^+$,
where $\theta_{\mbox{open}} <$ 3 mrad for an 80 GeV neutron.  OPE model
descriptions of neutron production do not in general predict the backward multiplicity.

\begin{table}
\begin{tabular}{lcc}
\hline
          & forward                      & backward \\
\hline
neutron   & $-0.090 \pm 0.006 \pm 0.009$ & $ 0.003 \pm 0.004 \pm 0.003$ \\
photon    & $-0.009 \pm 0.015 \pm 0.007$ & $-0.019 \pm 0.010 \pm 0.003$ \\
$\pi^0$   & $-0.022 \pm 0.030 \pm 0.002$ & $ 0.007 \pm 0.021 \pm 0.001$ \\
\hline
\end{tabular}
\caption{Asymmetries measured by the EMCal.
The errors are statistical and systematic, respectively.
There is an additional scale uncertainty, due to the beam polarization uncertainty, of $(1.0^{+0.47}_{-0.24})$.}
\label{tbl:asym}
\end{table}


%

In summary, we have measured the single transverse-spin 
asymmetry, $A_N$, for 
photon, $\pi^0$, and neutron production at very forward and very backward
angles in $p\!\!\uparrow$+$p$ 
collisions at $\sqrt{s}$ = 200~GeV. The 
asymmetries for photon and $\pi^0$ were consistent with 
zero, whereas a significant asymmetry
was observed for forward neutrons.
The large neutron asymmetry
is accompanied by a small forward multiplicity and a large backward
multiplicity.
The asymmetry for neutrons produced backward from the polarized beam
was consistent with zero.
This newly discovered asymmetry can be used for a non-destructive 
polarimeter to monitor the beam polarization in collision
and the physics can be explored with a dedicated hadronic
calorimeter.

%


We thank the staff of the RHIC project, Collider-Accelerator, and Physics
Departments at BNL and the staff of PHENIX participating institutions for
their vital contributions.  We thank the staff of SLAC and the SLAC facility for support and use of the Final Focus Test Beam.
We also thank S. White of Brookhaven for loan on short notice of the ZDC module.
We acknowledge support from the Department of
Energy and NSF (U.S.A.), MEXT and JSPS (Japan).


\begin{thebibliography}{0}

\bibitem{E704} D.~L.~Adams {\it et al.} [FNAL E581/704 Collaboration], Phys.\ Lett.\ B {\bf 261}, 201 (1991).
\bibitem{OPE} M.~Bishari, Phys.\ Lett.\ B {\bf 38}, 510 (1972).
\bibitem{Flauger76} W.~Flauger, F.~M\"{o}nnig, Nucl.\ Phys.\ B {\bf 109}, 347 (1976).
\bibitem{Blobel78} V.~Blobel {\it et al.}, Nucl.\ Phys.\ B {\bf 135}, 379 (1978).
\bibitem{D'Alesio99} U.~D'Alesio, H.~J.~Pirner, Eur.\ Phys.\ J.\ A {\bf 7}, 109 (2000).
\bibitem{Berger75} E.~L.~Berger, A.~C.~Irving, Phys.\ Rev.\ D {\bf 12}, 3444 (1975).
\bibitem{Kopeliovich96} B.~Kopeliovich, B.~Povh, I.~Potashnikova, Z.\ Phys.\ C {\bf 73}, 125 (1996).
\bibitem{Soffer} J.~Soffer, N.~A.~T\"{o}rnqvist, Phys.\ Rev.\ Lett.\ {\bf 68}, 907 (1992).
\bibitem{Bunce} G.~Bunce {\it et al.}, Phys.\ Rev.\ Lett.\ {\bf 36}, 1113 (1976).
\bibitem{regge} P.~D.~B.~Collins, Introduction to Regge Theory, Cambridge University Press (1977).
\bibitem{jinnouchi} O.~Jinnouchi {\it et al.}, AIP Conf.\ Proc.\ {\bf 675}, 817 (2003) [Czech.\ J.\ Phys.\ {\bf 53}, B409 (2003)].
\bibitem{Tojo} J. Tojo {\it et al.}, Phys. Rev. Lett. {\bf 89} (2002) 052302.
\bibitem{trueman} T.~L.~Trueman, arXiv:hep-ph/0203013.
\bibitem{GEANT} R. Brun {\it et al.}, GEANT 3 Manual, CERN Program Library Long Writeup W5013, 1994.
\bibitem{spinka} G.~G.~Ohlsen, P.~W.~Keaton, Nucl.\ Instrum.\ Meth.\ {\bf 109}, 41 (1973);
H. Spinka, Argonne National Laboratory Report No. ANL-HEP-TR-99-113, 1999.
\bibitem{ZDC} C.~Adler, A.~Denisov, E.~Garcia, M.~J.~Murray, H.~Strobele, S.~White, Nucl.\ Instrum.\ Meth.\ A {\bf 470}, 488 (2001).
\bibitem{Togawa} M.~Togawa, to appear in the proceedings of 17th International Spin Physics Symposium (SPIN06), Kyoto, Japan, 2-7 Oct 2006.

\end{thebibliography}

\end{document}